# Some peculiarities of activity for comets with orbits on 2 - 5 AU

*E. Yu. Musiichuk[1], S. A. Borysenko[2],*

[1]Taras Shevchenko National University of Kyiv, Glushkova ave., 4, 03127, Kyiv, Ukraine
[2]Main astronomical observatory, National Academy of Sciences, 27 Akad. Zabolotnoho Str., 03143, Kyiv, Ukraine
borisenk@mao.kiev.ua

Periodic comets of different dynamical groups with orbits at 2 - 5 AU still occasionally active. The observed dust activity of such objects can be connected with processes of water ice sublimation (MBCs) or crystallization of amorphous water ice (QHCs) as well as with external causes. No connections of cometary flares with cyclic variations of solar activity indexes were found. But some individual solar flares can affect the brightness of comets. Cometary objects in the main asteroid belt have lower statistic of flares than comets at orbits like to quasi-Hilda objects.

Key words: Sun: activity, flares; comets: general

## Introduction

Main-belt comets (MBCs) are bodies orbiting within the asteroid belt that have shown comet-like activity during part of their orbit. The first active object in the main-belt was discovered in 1996. About two dozens of main-belt comets (MBCs) were discovered for the last 20 years. Many of them were discovered at outburst [3]. We present heliocentric distribution of cometary flares as a function of heliocentric longitude and heliocentric distance. Also we used the same analysis for some ecliptic [1] and quasi-Hilda comets (QHCs) as a comparison. A quasi-Hilda comet (QHC) is a Jupiter-family comet that interacts strongly with Jupiter and undergoes extended temporary capture by it. These comets are associated with the Hilda asteroid zone in the 3:2 inner mean-motion resonance with Jupiter. Typically, asteroids in this zone have a semimajor axis between 3.70 and 4.20 AU, eccentricities below 0.30, and inclinations of no more than 20 degrees [11].

Usually, comets with orbits at the heliocentric distances more than 2 AU don't have emissions in their spectrum. Only dust activity can be observed occasionally in the MBCs as well as in the most of QHCs. In this reason, it is interesting to compare the frequency of cometary flares with the indices of the 11-year cyclic activity of the Sun.

Some orbital and physical parameters of the selected comets are listed in Table 1.

## Heliocentric distribution of cometary flares

We used data of cometary brightness from on-line databases: Minor Planet Center (minorplanetcenter.net) and Seiichi Yoshida's database (www.aerith.net) Heliocentric longitude at the maximum of outburst was calculated using JPL's HORIZONS system (ssd.jpl.nasa.gov) (Fig.1). We chose only cometary flares with amplitudes $2^m$ and more to avoid variations of brightness caused by rotation or photometric errors.

We used data of Solar X-ray fluxes (1 - 8 Å) from SPACE WEATHER PREDICTION CENTER (SWPC)(https://www.swpc.noaa.gov) by results of Geostationary Operational Environmental Satellite sys-tem (GOES). The GOES operated by the United States' National Oceanic and Atmospheric Administration's National Environmental Satellite, Data, and Information Service division, supports weather fore-casting, severe storm tracking, and meteorology research. Spacecraft and ground-based elements of the system are work together to provide a continuous stream of environmental data (http://www.polarlicht-vorhersage.de/goes_archive).

Large solar X-ray flares can change the Earth's ionosphere, which blocks high-frequency (HF) radio transmissions on the sunlit side of the Earth. Solar flares are also associated with Coronal Mass Ejections (CMEs) which can ultimately lead to geomagnetic storms. SWPC sends out space weather alerts at the M5 level. Some large flares are accompanied by strong radio bursts that may interfere with other radio frequencies and cause problems for satellite communication and radio navigation (GPS). The intensive solar X-rays also can be reason of cometary X-ray emission [9].

Table 1: Orbital and physical parameters of selected comets

| Comet | $q$ (AU) | $a$ (AU) | $e$ | $i$ (°) | P (yr) | $T_J$ | $H_{10}(nucl)$ | Type |
|---|---|---|---|---|---|---|---|---|
| 47P/Ashbrook – Jackson | 2.82 | 4.13 | 0.32 | 13.03 | 8.38 | 2.907 | 13.6 | JFC |
| 65P/Gunn | 2.91 | 3.89 | 0.25 | 9.18 | 7.67 | 2.999 | 12.0 | QHC |
| 74P/Smirnova – Chernyh | 3.54 | 4.16 | 0.15 | 6.65 | 8.49 | 3.007 | 12.1 | QHC |
| 91P/Russell 3 | 2.61 | 3.89 | 0.33 | 14.09 | 7.69 | 2.920 | – | JFC |
| 116P/Wild 4 | 2.19 | 3.49 | 0.37 | 3.61 | 6.51 | 3.009 | 13.0 | QHC |
| 117P/Helin – Roman – Alu 1 | 3.06 | 4.10 | 0.25 | 8.70 | 8.29 | 2.967 | 11.4 | QHC |
| 129P/Shoemaker – Levy | 3.91 | 4.29 | 0.09 | 3.44 | 8.89 | 3.019 | – | QHC |
| 133P/Elst – Pizarro | 2.66 | 3.17 | 0.16 | 1.39 | 5.63 | 3.183 | 15.7 | MBC |
| 152P/Helin – Lawrence | 3.10 | 4.49 | 0.31 | 9.88 | 9.50 | 2.900 | – | JFC |
| 176P/LINEAR | 2.58 | 3.20 | 0.19 | 0.23 | 5.71 | 3.166 | 15.2 | MBC |
| 180P/NEAT | 2.49 | 3.86 | 0.36 | 16.87 | 7.58 | 2.889 | 14.3 | JFC |
| 187P/LINEAR | 3.88 | 4.60 | 0.16 | 13.59 | 9.87 | 2.936 | 13.9 | QHC |
| 202P/Scotti | 2.53 | 3.79 | 0.33 | 2.19 | 7.37 | 2.982 | 15.7 | QHC |
| 211P/Hill | 2.35 | 3.56 | 0.34 | 18.89 | 6.70 | 2.934 | – | QHC |
| 219P/LINEAR | 2.37 | 3.65 | 0.35 | 11.53 | 6.98 | 2.961 | 14.9 | QHC |
| 232P/Hill | 2.98 | 4.49 | 0.33 | 14.63 | 9.51 | 2.853 | 14.5 | JFC |
| 238P/Read | 2.37 | 3.17 | 0.25 | 1.26 | 5.63 | 3.153 | 17.6 | MBC |
| 243P/NEAT | 2.45 | 3.83 | 0.36 | 7.64 | 7.50 | 2.946 | 15.0 | QHC |
| 244P/Scotti | 3.93 | 4.90 | 0.20 | 2.26 | 10.8 | 2.963 | 12.5 | QHC |
| 246P/2010 V2 (NEAT) | 2.88 | 4.03 | 0.29 | 15.97 | 8.08 | 2.914 | 11.3 | QHC |
| 266P/Christensen | 2.33 | 3.53 | 0.34 | 3.43 | 6.64 | 3.020 | – | QHC |
| 287P/Christensen | 3.05 | 4.18 | 0.27 | 16.30 | 8.54 | 2.902 | 14.8 | QHC |
| 288P/(300163) | 2.44 | 3.05 | 0.20 | 3.24 | 5.32 | 3.204 | 16.2 | MBC |
| 297P/Beshore | 2.34 | 3.45 | 0.32 | 10.33 | 6.40 | 3.027 | 13.8 | QHC |
| 324P/La Sagra | 2.62 | 3.10 | 0.15 | 21.42 | 5.45 | 3.100 | 15.5 | MBC |
| 331P/Gibbs | 2.88 | 3.00 | 0.04 | 9.74 | 5.20 | 3.229 | 15.4 | MBC |
| 340P/Boattini | 3.06 | 4.25 | 0.28 | 2.08 | 8.76 | 2.959 | 14.7 | QHC |
| 345P/LINEAR | 3.16 | 4.05 | 0.22 | 2.72 | 8.14 | 3.005 | 15.2 | QHC |
| 348P/PANSTARRS | 2.21 | 3.17 | 0.30 | 17.57 | 5.64 | 3.061 | 18.0 | MBC |
| 354P/LINEAR | 2.00 | 2.29 | 0.12 | 5.25 | 3.47 | 3.583 | – | MBC |
| 358P/PANSTARRS | 2.40 | 3.15 | 0.24 | 11.06 | 5.59 | 3.135 | 17.2 | MBC |
| P/1999 XN120 (CATALINA) | 3.30 | 4.18 | 0.21 | 5.03 | 8.55 | 2.990 | 13.7 | QHC |
| P/2010 H2 (Vales) | 3.10 | 3.85 | 0.20 | 14.27 | 7.55 | 2.987 | – | QHC |
| P/2013 R3 (Catalina – PANSTARRS) | 2.21 | 3.04 | 0.27 | 0.90 | 5.29 | 3.184 | 16.0 | MBC |
| P/2015 F1 (PANSTARRS) | 2.54 | 3.52 | 0.28 | 2.80 | 6.62 | 3.055 | – | QHC |

Distribution of flares for the main-belt comets are strongly inhomogeneous (Fig. 1 (left)). It is not easy to explain this only by law statistical sampling. Most of outbursts occurred for comets at heliocentric ecliptic longitudes 0 – 60 degrees. All flares in this zone are located along of some arc, which can be an arc of hypothetically orbit of meteor stream. Such stream can not be observable at the Earth due to high eccentricity of the hypothetical orbit ($e > 0.5$) and perihelion distance $q > 1$ AU with low inclination ($i < 10$).

It is not clear is the power of meteor stream sufficient to activate comet and increase its brightness. Probably, only concentration of big fragments in the central part of such hypothetical stream can have possibility to disturb dust layer on surface of the main-belt object [2].

For quasi-Hilda comets and some Jupiter-family comets with close orbits we have more homogeneous distribution of flares with good statistic (Fig. 1 (right)).

Time intervals of flares between GOES registrations and observations of cometary activity in most cases less than 20 days for QHCs, and about 10 days for MBCs (Fig. 2).

Similar analysis for high-energy proton flux data (> 10 MeV) show some smaller number of connected cometary flares. In this case, time interval of "cometary reaction" is only about few days (Fig. 3).

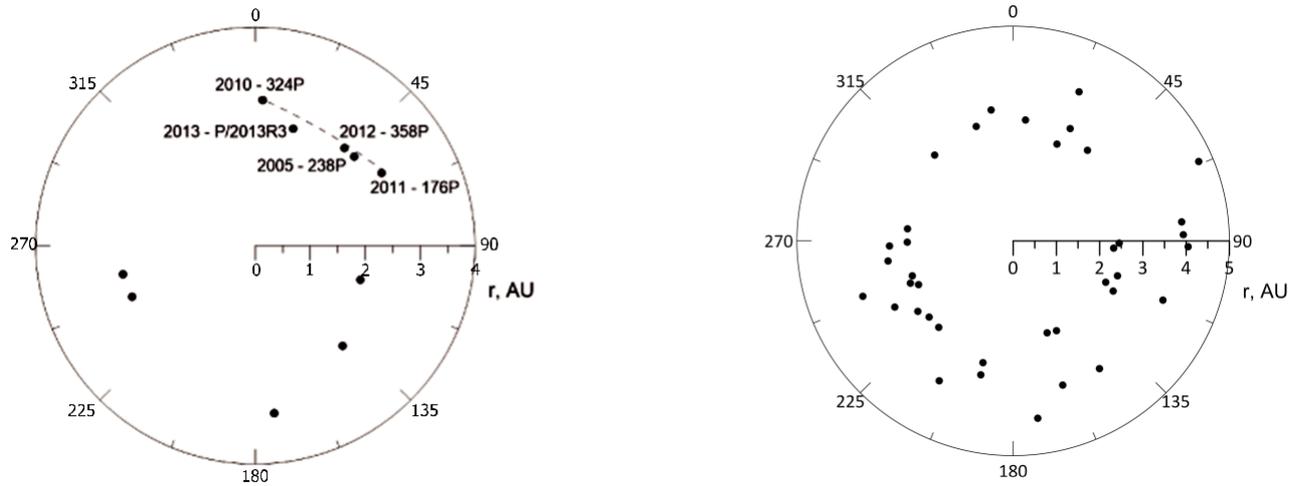

Fig. 1: Heliocentric distribution of cometary flares for MBCs (left) and QHCs (right) over the past 20 years.

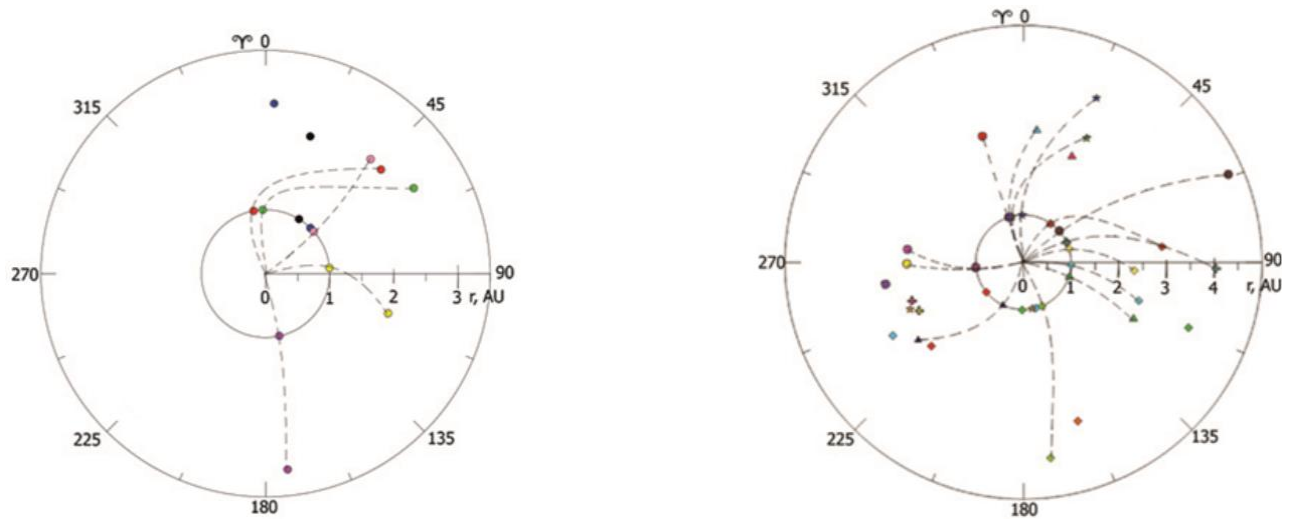

Fig. 2: Mutual heliocentric distribution of registered solar X-ray flares and cometary flares for MBCs (left) and QHCs (right). (Cometary flares, which can be connected with the solar flares, are signed by the dashed line).

For QHCs only about half of flares can be explained by solar activity, but only several for MBCs (Fig. 2). In any case, the analysis of these statistical distributions requires more observational data.

## Correlations of cometary flares with solar activity

We use values indexes of solar activity for the last 20 years to determinate of their correlation with flares for different dynamical groups of comets: SSN – solar spots number; F10.7 – radioflux on the wavelength 10.7 cm; $L_\alpha$ – ultraviolet flux of hydrogen line; $H_\alpha$ – infrared flux of hydrogen line. All of the indexes are directly connected with chromospheric activity of the Sun.

A correlation coefficient was calculated as:

$$k = \frac{\sum_{i=1}^{n}(X_i - \bar{X})(Y_i - \bar{Y})}{\sqrt{\sum_{i=1}^{n}(X_i - \bar{X})^2(Y_i - \bar{Y})^2}}$$

where $X_i$ – quantity of solar activity index per year, $Y_i$ – quantity of cometary flares per year,

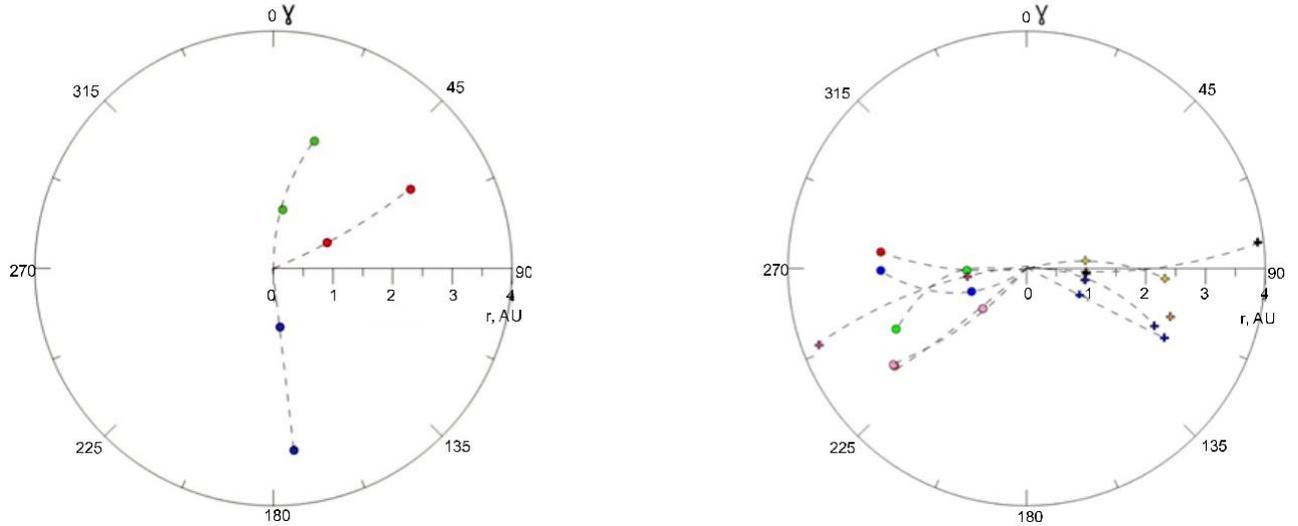

Fig. 3: Mutual heliocentric distribution of registered solar proton fluxes and cometary flares for MBCs (left) and QHCs (right). (Cometary flares, which can be connected with the solar proton fluxes are signed by the dashed line).

$$\bar{X} = \frac{1}{n}\sum_{i=1}^{n} X_i, \quad \bar{Y} = \frac{1}{n}\sum_{i=1}^{n} Y_i$$

Table 2: Correlation coefficients of the cometary flares with solar activity indices

| Index | k(QHCs) | k(MBCs) |
|---|---|---|
| SSN | -0.16 | -0.37 |
| F10.7 | -0.17 | -0.32 |
| L | -0.12 | -0.35 |
| H | -0.24 | -0.33 |

Received correlation coefficients $|k| < 0.4$ not show any significant dependence between the indexes of solar activity and flares for MBCs as for others ecliptic comets with orbits like to QHCs.

## Heliocentric distribution of solar radiation

The level of irradiance for X-ray solar fluxes is about $10^{-4}$ $W/m^2$ (for M5-class flares) and more (for X-class flares) accordingly to registered satellite data. The background values for the irradiance of X-ray solar fluxes in the phase of the minimum solar activity, usually, do not exceed the values ($10^{-6} \div 10^{-7}$ $W/m^2$).

Generally, solar radiation decreases inversely to the square of heliocentric distance

$$f = \frac{f_\oplus r_\oplus^2}{r^2}$$

where: $f_\oplus$ is a solar constant at the Earth surface, $r_\oplus$ is the distance from Earth to the Sun and $r$ is the distance from the Sun to the object in meters.

The table below gives standardized values for the radiation at each of some selected bodies.

Table 3: Values for the radiation at some bodies

| Object | $r$ ($10^{11}$ m) | $f$ ($W/m^2$) |
|---|---|---|
| Earth | 1.496 | 1367 |
| MBCs | 4.488 | 152 |
| QHCs | 5.984 | 85 |

The solar constant is an average of a varying value. Over the last three 11-year sunspot cycles it has varied by approximately 0.1 percent (about ±1 $W/m^2$). As one can see from Table 3, the solar radiation at QHCs is more than an order less as for the Earth.

## Discussion and conclusions

For a majority of objects, the observational, compositional, and dynamical definitions of cometary activity lie in close agreement. Some of comets, for example, 65P, 74P, 246P can be observable along the whole their orbit (with exception of Sun's conjunctions). Occasionally such comets increase their brightness more then by 2 mag. Their behavior likes to 29P comet with orbit on about 6 AU. Sometime, the activity of comets at large heliocentric distances more than 3 AU cannot be explained in the frame of the standard model [12], in which the sublimation of water ice heated by solar radiation is the primary cause of the activity of the comet nucleus. [5]. Some different sources of energy required for the dust and gas release at large heliocentric distances, such as meteoritic impacts, crystallization of amorphous water ice and polymerization of hydrogen cyanide [2].

At heliocentric distances larger than 3 AU different processes, with respect to water sublimation, must be invoked to drive the presence of the coma, for example *CO* or *$CO_2$* ice direct sublimation or gas release. The dust environment driven by one of these phenomena may differ from those caused by water sublimation, and may give hints on the nature, composition and physical characteristics of the comet [8, 10].

The sublimation of water ice is one of the significant reasons for activity of comets with orbits at the outer edge of the main asteroid belt. Some of such objects can eject dust, unexpectedly producing transient, comet-like comae and tails. Observational evidence for the sublimation of water ice is strongest in the two repeatedly active objects 133P/Elst – Pizarro and 238P/Read. Activity in 176P can be interpreted in terms of a sublimation model that is able to fit the observed coma morphology. The same must be said for observations of P/2008 R1, where sublimation can fit the data but where an impact origin may also apply [4, 6, 7]. At the same time, cometary flares of comets 176P, 238P, 258P are in agreement with some solar X-ray flares (Fig. 1).

Direct correlation between the variations of solar activity and cometary flares was not found (see Tab.2). Nevertheless, some cometary flares can be connected with individual active processes on the Sun. Therefore, comets, as possible indicators of solar activity, can be used to resolve some fundamental and applied problems.